\documentclass[twocolumn,aps,prl, superscriptaddress,showpacs]{revtex4-1}
\usepackage[latin9]{inputenc}
\usepackage{amssymb}
\usepackage{amsmath}
\usepackage{color}
\usepackage[export]{adjustbox}
\makeatletter

\usepackage{graphicx}
\usepackage{epstopdf}

\makeatother

\begin{document}

\title{Weak localization competes with the quantum oscillations in a natural electronic superlattice: the case of Na$_{1.5}$(PO$_{2}$)$_{4}$(WO$_{3}$)$_{20}$}

\author{Kamil K. Kolincio} 
\email{kamkolin@pg.edu.pl}
\affiliation{Laboratoire CRISMAT, UMR 6508 du CNRS et de l'Ensicaen, 6 Bd Marechal
Juin, 14050 Caen, France.}
\affiliation{Faculty of Applied Physics and Mathematics, Gdansk University of Technology,
Narutowicza 11/12, 80-233 Gdansk, Poland}

\author{Olivier P\'erez}
\affiliation{Laboratoire CRISMAT, UMR 6508 du CNRS et de l'Ensicaen, 6 Bd Marechal
Juin, 14050 Caen, France.}

\author{Enric Canadell}
\affiliation{Institut de Ci\`encia de Materials de Barcelona (ICMAB-CSIC), Campus Bellaterra, 08193 Bellaterra, Barcelona, Spain}

\author{Pere Alemany}
\affiliation{Departament de Ci\`encia de Materials i Qu\'imica Fisica and Institut de Qu\'imica Te\`orica i Computacional (IQTCUB), Universitat de Barcelona, Mart\'i i Franqu\`es 1, 08028 Barcelona, Spain }

\author{Elen Duverger-N\'edellec}
\affiliation{Laboratoire CRISMAT, UMR 6508 du CNRS et de l'Ensicaen, 6 Bd Marechal
Juin, 14050 Caen, France.}
\affiliation{Charles University, Faculty of Mathematics and Physics, Condensed Matter Department, Ke Karlovu 3, 121 16 Praha 2, Czech Republic}

\author{Arianna Minelli}
\affiliation{ESRF - The European Synchrotron, 71, Avenue des Martyrs, F-38000 Grenoble, France}

\author{Alexei Bosak}
\affiliation{ESRF - The European Synchrotron, 71, Avenue des Martyrs, F-38000 Grenoble, France}

\author{Alain Pautrat}
\affiliation{Laboratoire CRISMAT, UMR 6508 du CNRS et de l'Ensicaen, 6 Bd Marechal
Juin, 14050 Caen, France.}

\begin{abstract}
We report an investigation of the combined structural and electronic properties of the bronze Na$_{1.5}$(PO$_{2}$)$_{4}$(WO$_{3}$)$_{20}$.   Its low dimensional structure and possible large reconstruction of the Fermi surface due to charge density wave instability makes this bulk material a natural superlattice with a reduced number of carriers and Fermi energy.  Signatures of multilayered 2D electron weak localization are consequently reported, with an enhanced influence of quantum oscillations. A crossover between these two antagonistic entities previously observed only in genuine low dimensional materials and devices, is shown to occur in a bulk crystal due to its hidden 2D nature.

\end{abstract}

\maketitle

The unfading interest in low dimensional conducting materials originates from their remarkable electronic properties. High critical temperature in superconductors \cite{Kharkov2016, Cyr2017},
improved thermoelectric properties \cite{Xie2009, Mao2016, Kolincio2016_1, Molina2018}, 
instability of the electron gas into a charge density wave (CDW) \cite{Monceau2012, Gruner1988}, are consequences of a pronounced structural and electronic anisotropy. In particular, quasi 1D or 2D conductors exhibit a partial Fermi surface (FS) reconstruction at a CDW transition, accompanied by strong reduction of free electron density, increase of the remaining carriers  mobilities, and enhanced magnetoresistance \cite{Yasuzuka1999,Schlenker1985}. 

In addition, quantum
corrections may strongly impact the low temperature transport properties.
Weak localization (WL) phenomenon is a well known example. Another characteristic
of low temperature magnetoresistance effects in metals, especially of those with high carrier mobility, is the existence
of Shubnikov de Haas (SdH) quantum oscillations. They originate from
the quantization of electron density of states into Landau levels
when exposed to a magnetic field, and are observed in high purity
samples in the limit $\omega_c\tau\gg$ 1, where $\omega_c=\frac{eB}{m^{*}}$ is the
cyclotron frequency, $m^{*}$ the effective mass of carriers, and
$\tau$ the elastic scattering time). WL and SdH effects are a priori antagonistic
because they are respectively favored and limited by disorder. However,
due to the decrease of the WL contribution with magnetic field and
with the restricted condition $\omega_c\tau\leq(k_{F}\ell)^{-1}$ ($k_{F}$ is the Fermi
momentum and $l$ the electronic mean free path), a crossover
can be observed between these two quantum processes, whose magnetic field range
has been related to interaction effects \cite{Sedrakyan2008}. This
crossover has been previously observed in rare systems showing strong
2D characteristics: 2D electron gas heterostructures with high mobilities \cite{Wang2018},
quantum wells \cite{Sedrakyan2008, Tayari2015, Sarcan2018}, and graphene \cite{Heer2010, Jabakhanji2014}. We
show here that both WL and SdH are for the first time observed in a single
crystal whose 2D character arises from the carriers confinement in conducting
layers with a thickness of few atomic planes. The sample under study belongs to
the monophosphate tungsten bronze (MPTB) family A$_{x}$(PO$_{4}$)$_{2}$(WO$_{3}$)$_{2m}$ (A = K, Na or Pb).
In this family, carriers originate from the PO$_{4}$ groups, and
are delocalized in the middle of WO$_{3}$ layers, giving a
pronounced anisotropic electronic structure \cite{Roussel2001, Foury2002}. The carrier
density can be tuned by changing $m$ due to the increase of unit cell
for constant carrier numbers and/or by the insertion of A cations
with metallic character \cite{Roussel2000}. These A cations are inserted
in hexagonal tunnels to give A$_{x}$(PO$_{4}$)$_{2}$(WO$_{3}$)$_{2m}$\cite{Domenges1988}.
Supplemental interest of this bronze family is that most of the investigated
members have shown CDW transitions that have been explained
using the hidden nesting concept \cite{Canadell1991, Kolincio2016}.

\begin{figure*}[t!]
\includegraphics[width=0.9\textwidth]{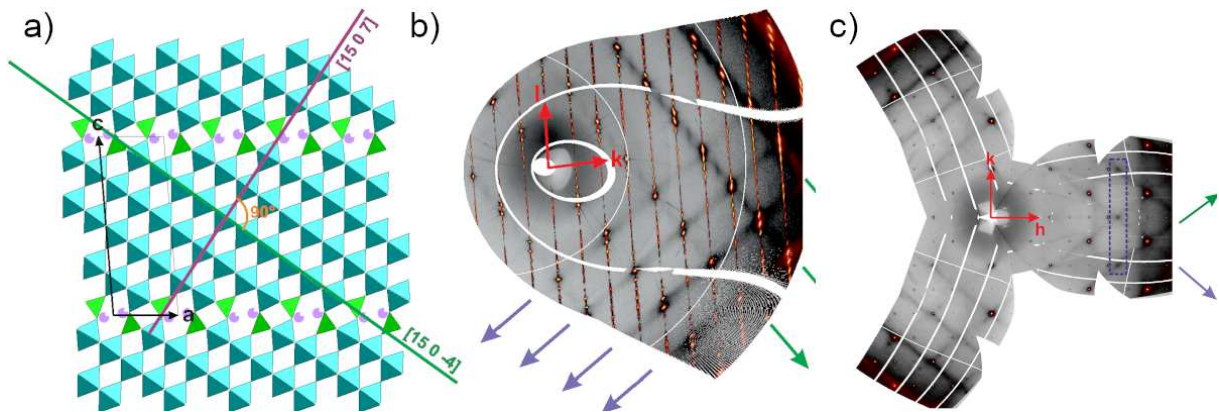}
 \caption{\label{str} a) Projection along \textbf{b} of the structure of $Na_{1.5}P_4 W_{20} O_{68}$. Blue octahedra are ($WO_6$), green tetrahedra are ($PO_4$) and purple a circles are Na atoms. The [15 0 7] and [15 0 -4] directions are showing the short and long direction ($WO_6$) chains; they are 90$^{\circ}$ oriented. b) (0kl)* c) (hk0)* planes assembled from the frames collected at 130 K on the ID28 beamline at ESRF\cite{Girard2019}. Two directions of diffuse scattering are shown by green and blue arrows. Intensifications of the diffuse signal are observed at the intersection of the two types of diffusion scattering for $\frac{1}{2}$\textbf{a}* (see blue dashed rectangle).}
 \end{figure*}

X-ray diffraction investigations performed at room temperature reveal that the whole pattern can be described using a monoclinic cell  ($a = 6.5483(6)$\AA, $b= 5.2893(5)$\AA, $c= 18.0789(15)$\AA, $\beta = 94.7000(4)^{\circ}$) and the space group $P2_1/m$; these cell parameters are in agreement 
with those reported for Pb$_x$P$_4$W$_{20}$O$_{68}$ \ on the basis of a pattern matching analysis performed on powder diffraction data~\cite{roussel1998}. 
The final agreement factor for our data is  4.7\% and the refined chemical composition is
 Na$_{1.49(8)}$P$_4$W$_{20}$O$_{68}$; Na sites are partly occupied. For details of the data collection, refinement and the atomic structure parameters see the Supplementary Material \cite{supmat}. 

A projection along the \textbf{b} axis is shown in Figure \ref{str}a. The main structural characteristics expected for the MPTB family in the fundamental state are observed. Slabs of the (WO$_6$) edge sharing octahedra alternate with slices of (PO$_4$) tetraehdra; at the junction, hexagonal tunnels host Na atoms. Two orthogonal directions [15 0 7] and [15 0 -4] can be highlighted; they correspond to chains of 10 and 5 edge sharing (WO$_6$) octahedra respectively (see Figure ~\ref{str}a). 

To check the possibility of a CDW transition, full data collections were performed at RT and 130 K on the ID28 beamline at ESRF
synchrotron. There is no evidence for additional scattering signal at RT but at 130K diffuse scattering can be observed on different experimental frames. An accurate study of the (hk0)* plane (see Figures \ref{str}c) reveals an enhancement of the diffuse signal at the intersection of two diffuse planes for the $\frac{1}{2}$\textbf{a*} wave vector.  
An analysis of the shift between the positions of the diffuse scattering in subsequent (0kl)*, (1kl)*, (2kl)* reciprocal layers, reveals that the diffuse sheets intersecting them run along the direction [15 0 7] (see Supplementary Material \cite{supmat} for details).

  The planes running along [15 0 7] are perpendicular to [-7 0 15]*; but [-7 0 15]* and [15 0 -4] fortuitously correspond to the same  direction (see Supplementary Materials \cite{supmat}) the direction of the chain of 10 octahedra. Then the observation of the diffuse scattering can be correlated with an ordering within the longest (WO$_6$) chain but with a loss of order both along \textbf{b} and along [15 0 -4] \textit{i.e.} with a loss of order with the adjacent chains. The existence of diffuse scattering planes evidences the 1D ordering, possibly associated to CDW. Nevertheless, the absence of their condensation to sharp reflections shows that this ordering shows at T = 130 K a moderate coherence length, instead of establishing a long range order.

  \begin{figure*}[t!]
\includegraphics[width=1.0\textwidth]{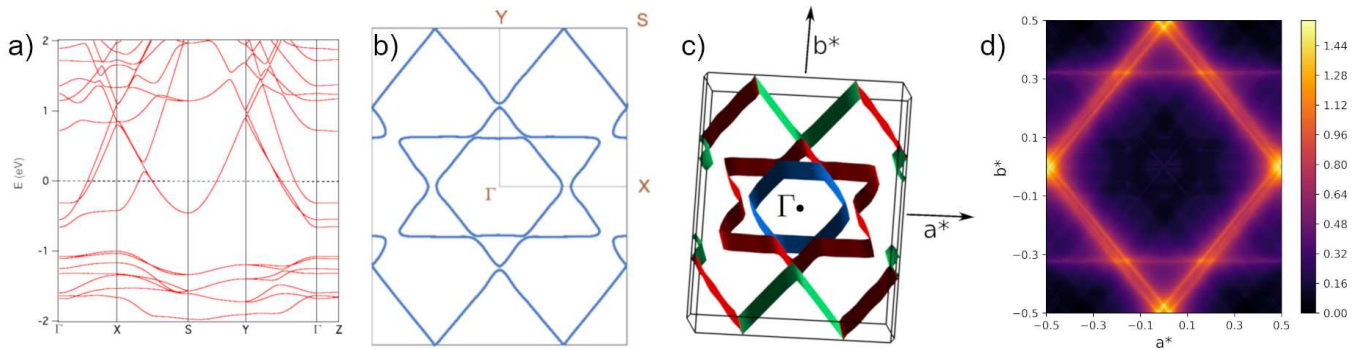}

\caption{\label{FS} a) Electronic band structure calculated for Na$_{1.5}$(PO$_{2}$)$_{4}$(WO$_{3}$)$_{20}$. Zero energy corresponds to the Fermi level and the different points are defined as $\Gamma$ = (0, 0, 0), X = (1/2, 0, 0), Y = (0, 1/2, 0), S = (1/2, 1/2, 0) and Z = (0, 0, 1/2), respectively in the units of the reciprocal vectors. b) section of the Fermi surface for $c^*=$ 0. c) 3D representation of the Fermi surface. d) calculated Lindhard function.}
\end{figure*}

Having established the crystalographic parameters, we studied the electronic structure of Na$_{1.5}$(PO$_{2}$)$_{4}$(WO$_{3}$)$_{20}$. The results of band calculations performed with Generalised Gradient Approximation (GGA) (see supplementary material for details \cite{supmat}) are shown in Fig. \ref{FS}. There are large similarities between the studied compound and other MPTB members showing CDW \cite{Wang1989, Canadell1991}. As seen in Fig. \ref{FS}a, three bands are crossing the Fermi level resulting in FS composed of three quasi 1D elements: three pairs of parallel planes running along $(a^*+b^*)$, $(a^* - b^*)$, and $b^*$ directions with almost no dispersion along $c^*$ as shown in Figures \ref{FS}a) and c). While the presence of similar flat regions in MPTB hosting CDW gives a rise to hidden 1D nesting \cite{Wang1989, Canadell1991}, their almost perfect alignment in Na$_{1.5}$(PO$_{2}$)$_{4}$(WO$_{3}$)$_{20}$ additionally enhances the susceptibility for such a scenario. The relevance of this mechanism is additionally supported by the calculation of a Linhard function (Fig. \ref{FS}d). In particular three very intense features are near the a*/2 wave vector suggest that large parts of the Fermi surface are susceptible to destruction by either an inherent instability of the Fermi surface or coupling with another type of instability of the lattice with this wave vector.

\begin{figure*}[t!]
  \includegraphics[width=1.0\textwidth]{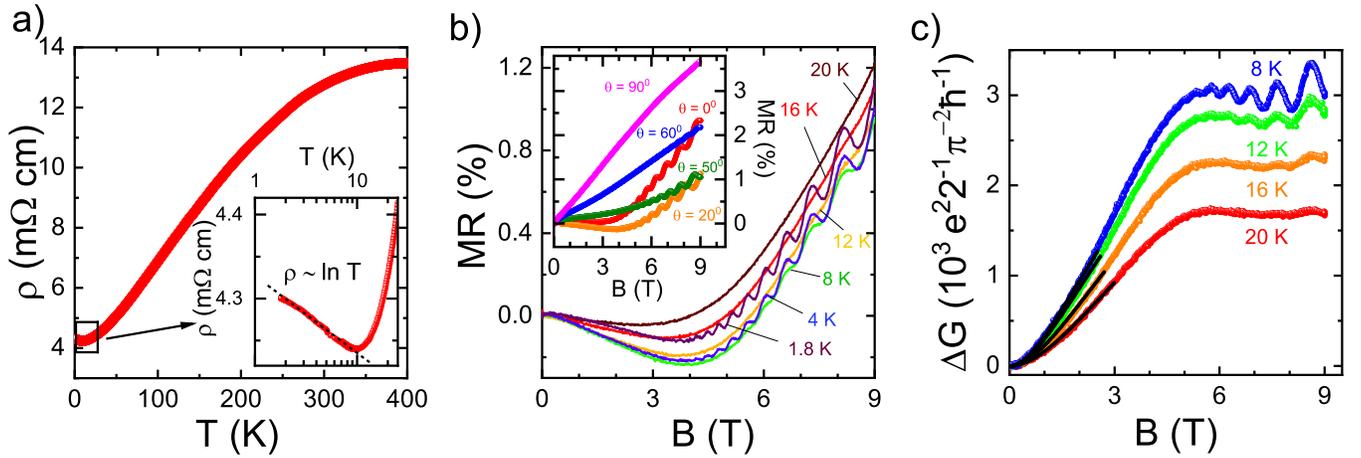}
 \caption{\label{transport} a) In - plane resistivity versus temperature. Inset: expanded view of the low temperature region shown with logarithmic horizontal scale. Dashed line is a guide for the eye. b) Magnetoresistance as a function of magnetic field applied parallel to c axis. Inset: MR(B) plots  at $T$ = 1.8 K for various $\theta$ angles between the magnetic field direction and the c axis. c) Magnetoconductance versus magnetic field for selected temperatures. The red curves show the data fits with equation (\ref{mreq}).}
 \end{figure*}

Fig. \ref{transport}a shows the thermal variation of resistivity $\rho$ of a Na$_{1.5}$(PO$_{2}$)$_{4}$(WO$_{3}$)$_{20}$
 single crystal. Metallic behavior ($\frac{d\rho}{dT}<0$) is observed
from 400 K down to 10 K, where a smooth increase of resistivity appears.
The electronic resistivity follows a -ln$T$
variation at low temperature, and this experimental observation
suggests a WL contribution.  The expression of the WL correction to conductance
for non interacting electrons is \cite{Bergmann1984}:

\begin{equation}
\Delta G = \frac{pe^{2}}{2\pi^{2}\hbar}ln\frac{T}{T_{0}}
\label{WL_res}
\end{equation}

where $p$ describes the temperature dependence of the inelastic scattering
rate $\tau_{i}\propto T^{-p}$. p is equal to 1 for phase breaking
dominated by Coulomb interactions and to 3 for electron-phonon scattering
\cite{Lee1985}. Fitting the low temperature part gives $\frac{dG}{dlnT}$
= 0.0517 $\Omega^{-1}$, to compare with $\frac{e^{2}p}{2\pi \hbar}$ = 1.23
10$^{-5}$ $\Omega^{-1}$ expected for a 2D WL contribution. The large
difference between these values can be explained if our sample is
considered as a superlattice of N independent and non-interacting 2D
layers, giving rise to a multilayered WL effect \cite{Moyle1987, Stormer1986, Szott1989, Szott1989_1, Szott_1992}. Dividing
the measured conductivity by the theoretical WL conductivity of a
single layer gives N $\approx$ 4300, considering a value
p = 1, as expected for electron scattering in d-electron systems. Multiplying N by a thickness of the W-O conducting layer - approximately equal to the c lattice constant, we obtain $\approx$ 10 $\mu$m, comparable to the thickness of the sample.
The weak disorder causing WL can originate from a weakly nonuniform distribution of Na atoms in the tunnels \cite{Kolincio2016}, stacking faults reported for MPTB with high $m$ parameter \cite{Domenges1983} or can be contributed by the short order of charge density waves state.

Since other mechanisms can lead to a logarithmic increase
of resistivity at low temperature (Kondo scattering, strong Coulomb
interaction), we analyzed the magnetoresistance ($MR = \frac{\rho(B)-\rho(0)}{\rho(0)}\cdot100\%$) in order to confirm the underlying mechanism.
The low temperature magnetoresistance depicted in Figure \ref{transport}b is non-monotonic. It is negative
for moderate $B$, and upon increasing the magnetic field, it shows a minimum followed by a notable increase. For T $\leq$ 18 K, large oscillations
of the resistance can be observed in the increasing part of $MR(B)$. 

\begin{figure}[b]
  \includegraphics[width=1.0\columnwidth]{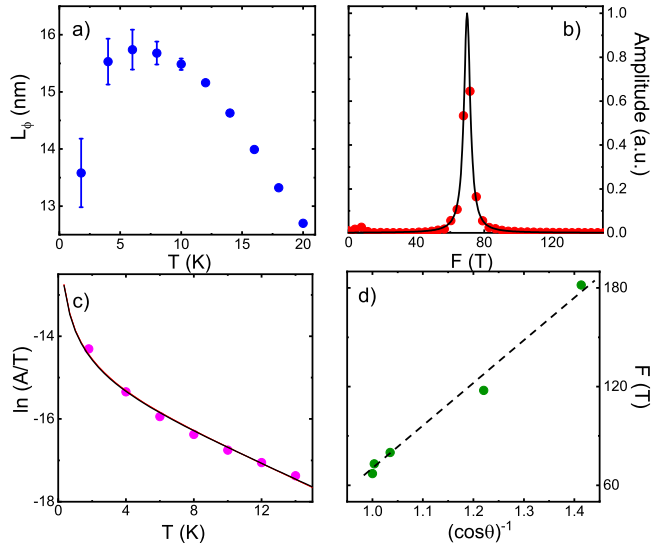}
 \caption{\label{fits} a) Temperature variation of the localization length, obtained from the magnetoconductance fits with equation \ref{mreq} b) Power spectrum of the oscillatory component of magnetoresistance. Solid line shows the Lorentzian fit to the experimental data. c) the geometrical dependence of the oscillation frequency. The dashed line is a guide for the eye.  d) Thermal variation of the oscillation amplitude.}
 \end{figure}

As a strong argument for WL, we observe a negative MR for B
perpendicular to the conducting plane ($\theta$ = 0$^o$) but it disappears
in the parallel configuration ($\theta$ = 90$^o$) - see inset of Fig. \ref{transport}b. This is precisely
what is expected in the 2D WL regime, where the orbital effect responsible
for the MR is suppressed for parallel fields \cite{Meyer_2002}. The low field magnetoconductance $\Delta G = G(B)-G(0)$  in the WL regime, can be described by the Hikami-Larkin-Nagaoka
 (HKN) expression \cite{Hikami_1980, Lee1985}:

\begin{equation}
\begin{split}
\Delta G =\frac{e^{2}}{2\pi^{2}\hbar}\left( \psi\left( \frac{1}{2}+\frac{1}{A}\right) +ln(A)\right),
\end{split}
\label{mreq}
\end{equation}

where $\psi$ is the digamma function and $A$ = $\frac{4eD\tau_{i}B}{\hbar}$.
To show the increase of magnetoconductance in the magnetic field, we have quantified and removed the classical $\sim$ B$^2$ magnetoresistance component. Then, we fit the low field curves with the WL expression considering the number of 4300 independent layers as found from the thermal dependence of conductivity. As shown in Fig. \ref{transport}c, the fits 
show good agreement with the experimental data. The diffusion length of localization
(Thouless length) L$_\phi$ = (D$\tau_{i}$)$^{1/2}$ was extracted from the
fitting parameter and reported in the Fig. \ref{fits}a. It is worth noting
that L$_\phi$ is larger than the thickness of  the conducting layer and approximately equal to the periodicity along the $c$ axis, confirming the regime of 2D coherence. When increasing the magnetic field, the dephasing
regime emerges for L$_{B}$ = $\left( \frac{2 \hbar}{eB_{\phi}}\right) ^{1/2}\geq L_{\phi}$.
The effect of the magnetic field is then no longer relevant for WL and a negative
contribution disappears. We have compared the value of $B_{\phi}$
extracted from the fitting with the condition L = L$_{B}$ and the magnetic
field value B$^{*}$ where the G(B) ceases to grow with B and where quantum
oscillations immediately emerge. We find that these values are very close to each other (for example for 4, 8  and 10 K the values obtained from the fits yield 5.27 T, 5.36 T, and 5.49 T respectively to compare with observed B$^*$ = 5.45 T, 5.31 T and 5.53 T, respectively). For lowest temperatures, $L_\phi$ ceases to grow as the temperature is lowered, and even drops for 1.8 K, where the negative MR term also starts to decrease. A plausible explanation for this effect is the further crossover from weak localization to weak antilocalization (WAL)\cite{Lang_2013, Liu_2012} due to strong spin-orbit effects observed in tungsten rich compounds \cite{Shenge_2017, McCormick_2017}. This scenario is also supported by the emergence of sharp cusp-like features in the low field limit of magnetoresistance (see Supplementary Materials), which is characteristic for WAL regime\cite{Bergmann1984, Thomas_2016}. A further examination of this term, to obtain reliable HKN fit with the introduction of the antilocalization component will require the transport measurements at temperatures below 1.8 K.

The oscillations in magnetoresistance emerge immediately after the negative WL term is suppressed, as evidenced in Fig. \ref{transport}c.  They are periodic as a function
of 1/B, as expected for Shubnikov-de Haas quantum oscillations. We
have used the Lifshitz-Kosevich (LK) theory to extract the electronic
parameters \cite{Lifshitz1955}. The Fourier transform of the oscillating part
of magnetoresistance gives the power spectrum shown in Fig. \ref{fits}b. A peak
can be observed at the frequency $F_{1}$= 69 T for $\theta$ = 0$^o$. This value yields the area $A_{1}$ of an extremal
Fermi surface cross section that is perpendicular to B, through the
Onsager relation \cite{shoenberg_1984} $A_{i}=2\pi eF_{i}/\hbar$. It corresponds to
a FS cross section $A_{1}$= 0.68 nm$^{-2}$, i.e. to a very small
part ($\approx$ 0.5 \%) of the first Brillouin zone, estimated to
be 110 nm$^{-2}$. A plausible explanation for the presence of such a small pocket is the Fermi surface decomposition associated with charge density wave. This scenario is supported by the X-ray diffuse scattering and electronic structure calculations showing a potential for strong nesting around a*/2, which most likely occurs at high temperature as for other high m values of the MPTB family \cite{Ottolenghi1996}.
The temperature dependence of the quantum oscillations amplitude is
depicted in Fig. \ref{fits}c. It can be related to the effective mass of quasi-particles
m$^{*}$ through the temperature damping factor. We find m$^{*}$=
0.83 m$_{e}$ with m$_{e}$ being the free electron mass, indicating that the electron-electron interactions are moderate as theoretically predicted for the WL-SdH crossover\cite{Sedrakyan2008}. The
field dependence of the quantum oscillations allows to estimate the
Dingle temperature T$_{D}$ $\approx$ 6 K, corresponding to an electronic
mean free path $l\approx$ 62 nm. Since the Fermi momentum is $k_{F}\approx(A_{1}/\pi)^{1/2}$ = 0.47
nm$^{-1}$, it leads to $k_{F}l\approx$ 29$\gg$ 1 confirming the diffusive regime \cite{Minkov2002, Pusep2004}.
The angle
$\theta$ between $B$ and the $c$ axis was varied from 0 to 90$^o$ and the SdH oscillations can be observed within experimental resolution up
to $\theta\approx$ 50$^o$. The frequency of the SdH oscillations
clearly shifts to larger values as $\theta$ is increased, with an
approximate 1/cos$\theta$ variation, as depicted in the inset of Figure \ref{fits}b. This is expected for quasi-2D
Fermi surface topology and is also consistent with the angular dependence of magnetoresistance shown in the inset of figure \ref{transport}b. For $B$ perpendicular to the conducting plane, MR initially attenuated by the negative contribution, steeply increases with $\sim B^2$ in the high field limit, where the WL phase coherence is broken, while for in-plane orientation MR is sublinear with $B$ which gives a tendency for saturation as $B \rightarrow \infty$. Such a behavior is characteristic for open and closed orbits, respectively, expected for a cylinder-like FS\cite{Pippard_2009}.

Such a reduced, cylindrical Fermi surface is a characteristic feature of 2D  systems, as high mobility ultrathin films and heterostructures \cite{Wu_2017, Gillgren_2014, Kozuka_2009, Li_2015} and unparalleled cases of pure Bi and graphene, where the FS covers only 10$^{-5}$ and 10$^{-6}$ of the Brillouin zone, respectively \cite{Soule_1964, Bhargava_1967, Behnia_2007, Edelman1976}. The renormalization of the Na$_{1.5}$(PO$_{2}$)$_{4}$(WO$_{3}$)$_{20}$ electronic structure driven by the charge density wave, causing the lowering of the Fermi wavevectors brings this bulk system close to a natural superlattice, with reduced number of carriers and/or reduced Fermi energy, which promotes the quantum phenomena.

In conclusion, we report the structural and transport properties
of the monophosphate tungsten bronze Na$_{1.5}$(PO$_{2}$)$_{4}$(WO$_{3}$)$_{20}$.
We demonstrate that this single crystal shows multilayered 2D weak localization effects.
As soon as the dephasing by magnetic field dominates and breaks the
WL contribution, there is a sudden crossover to the large quantum oscillations
regime. This shows that natural crystals with 2D characteristics and  reduced carriers density possibly due to a CDW condensation can be used as promising systems for investigating enhanced quantum transport properties.
\begin{acknowledgments}
Acknowledgements: A.P would like to thank Thierry Klein and Andrea
Gauzzi for valuable remarks on the competition between weak localization
and quantum oscillations conditions which motivated our analysis.
Financial support by the ANR Projects: ANR-18-CE92-0014 and  ANR-11-BS04-0004 is gratefully acknowledged. Work in Spain was supported by MICIU (PGC2018-096955-B-C44 and PGC2018-093863-B-C22), MINECO through the Severo Ochoa (SEV-2015-0496) and Maria de Maeztu (MDM-2017-0767) Programs and the Generalitat de Catalunya (2017SGR1506 and 2017SGR1289). E.Duverger-N\'edellec was supported by the project NanoCent-Nanomaterials center for advanced applications, project no. CZ.02.1.01/0.0/0.0/15\textunderscore 003/0000485, financed by the ERDF.
\end{acknowledgments}
\nocite{jana2006, superflip, Hohenberg1964, Kohn1965, Perdew1996,  Soler_2002, Artacho_2008, Troulier_1991, Kleinman1982, Artacho1999, Monkhorst1976, Poloni_2010}

%

\end{document}